\documentclass[aps,prb,reprint,amsmath,amssymb,nobibnotes]{revtex4-1}

\usepackage[ansinew]{inputenc}
\usepackage{graphicx}
\usepackage{textcomp}
\usepackage{bm}
\usepackage{units}
\usepackage{enumerate}
\input{glyphtounicode}
\newcommand{\ket}[1]{|#1\rangle}

\begin{document}

\title{Interference and dynamics of light from a distance-controlled atom pair\\ in an optical cavity}
\author{Andreas~Neuzner}
\author{Matthias~K\"{o}rber}
\author{Olivier~Morin}
\author{Stephan~Ritter}
\email[e-mail: ]{stephan.ritter@mpq.mpg.de}
\author{Gerhard~Rempe}
\affiliation{Max-Planck-Institut f\"{u}r Quantenoptik, Hans-Kopfermann-Strasse 1, 85748 Garching, Germany}

\maketitle
\textbf{Interference is central to quantum physics and occurs when indistinguishable paths exist, like in a double-slit experiment. Replacing the two slits with two single atoms\cite{eichmann_1993} introduces optical non-linearities for which nontrivial interference phenomena are predicted\cite{vogel_squeezing_1985,richter_g2int_1991,schoen_bunching_2001,skornia_bunching_2001,gruenwald_entanglement_2010}. Their observation, however, has been hampered by difficulties in preparing the required atomic distribution, controlling the optical phases and detecting the faint light. Here we overcome all of these experimental challenges by combining an optical lattice for atom localisation, an imaging system with single-site resolution, and an optical resonator for light steering. We observe resonator-induced saturation of resonance fluorescence\cite{alsing_suppression_1992,reimann_twoatoms_2015} for constructive interference of the scattered light and nonzero emission with huge photon bunching for destructive interference. The latter is explained by atomic saturation and photon pair generation\cite{richter_g2int_1991,schoen_bunching_2001,skornia_bunching_2001}. Our experimental setting is scalable and allows one to realize the Tavis-Cummings model\cite{tavis_cummings_1968} for any number of atoms and photons, explore fundamental aspects of light-matter interaction\cite{habibian_quantumlight_2011,macovei_2007,habibian_entanglement_2014,nikoghosyan_noon_2012,fernandez_nonlinear_2007}, and implement new quantum information processing protocols\cite{metz_quantumjumps_2006,pachos_ioncavgate_2002,kastoryano_dissent_2011,pellizzari_quantcomp_1994}.}

\begin{figure}
\includegraphics[width=88mm]{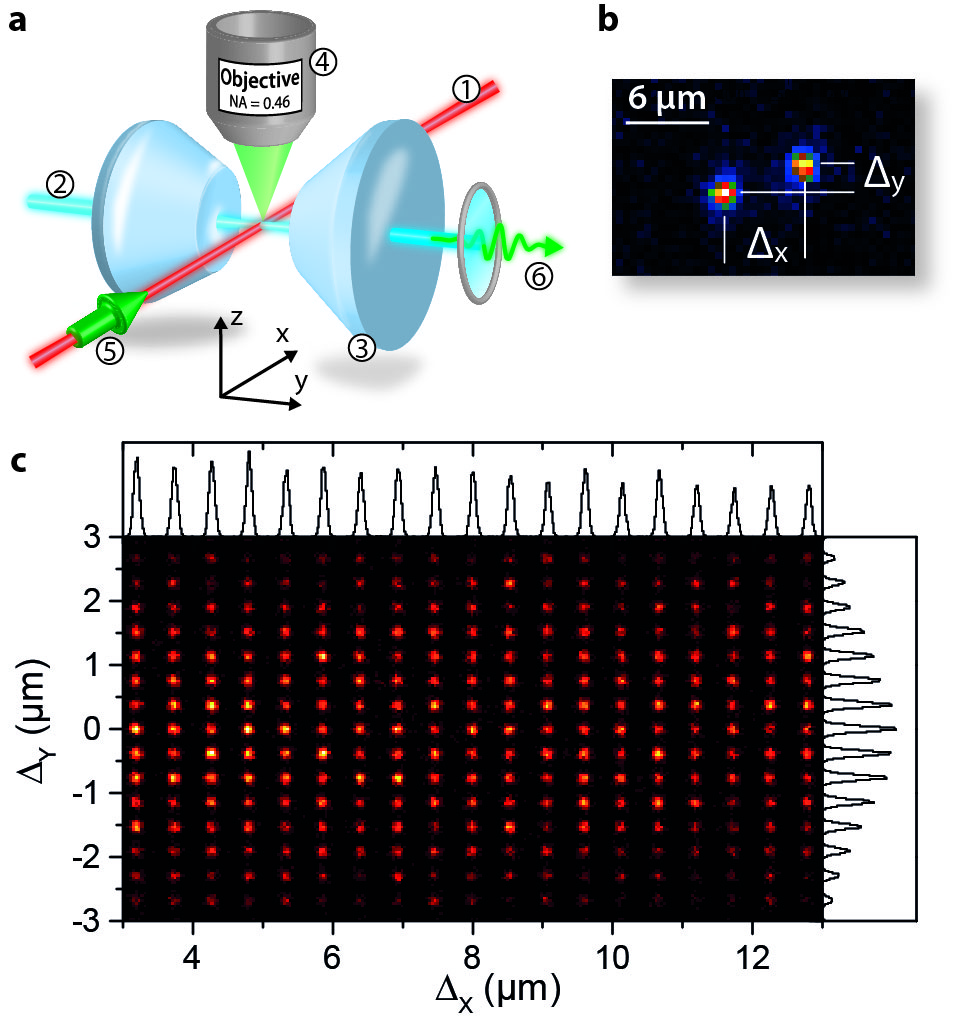}
\caption{\label{fig:setup}
\textbf{Experimental setup.} \textbf{a,} A two-dimensional optical lattice is formed by a retroreflected, red-detuned \textcircled{1} and a blue-detuned \textcircled{2} laser beam in a high-finesse optical resonator \textcircled{3}. A microscope objective \textcircled{4} is used to image and deterministically remove individual atoms trapped in the lattice. An atom pair is driven coherently with a running-wave beam \textcircled{5} propagating transversally through the resonator. Scattering from this beam into the single-sided cavity is studied via transmission through the outcoupling mirror \textcircled{6} as a function of the atomic positions. \textbf{b,} A typical fluorescence image of an atom pair, with atoms separated by several lattice sites along the x- and y-directions. In order to determine $\Delta_x$ and $\Delta_y$, Gaussian point spread functions are fitted to the individual atom images. \textbf{c,} A two-dimensional histogram of the difference coordinates between atom pairs reveals the lattice geometry. The data is corrected for a small non-orthogonality of the lattice. The one-dimensional projections onto the axes show that single-site resolution is achieved.}
\end{figure}

A multitude of non-classical radiation effects like photon antibunching\cite{carmichael_antibunching_1976} and squeezing\cite{walls_squeezing_1981} were predicted in the resonance fluorescence of single quantum emitters. The experimental observations of these effects\cite{kimble_antibunching_1977,atatuere_squeezing_2015} are milestones in the development of quantum optics. However, the large mismatch between the light mode driven by an emitter in free space and the light mode detected by an observer makes such quantum effects unpractical for applications. The way out is to couple the emitter to an optical resonator which enhances the interaction strength of the emitter with a tailor-made light mode. In fact, the experimental realization of well-controlled atom-cavity systems with single atoms as emitters has propelled the application potential of quantum-optical phenomena\cite{reiserer_rmp_2015} enormously and, in addition, has enabled the observation of fundamentally new quantum-mechanical radiation effects induced by the cavity\cite{birnbaum_blockade_2005,kubanek_twophoton_2008,ourjoumtsev_squeezing_2011}. When scaling these single-atom systems to multiple atoms, relative optical phases appear as a new degree of freedom. As those are determined by the spatial arrangement of the atoms, both, subwavelength localisation and precise knowledge about the relative positions of the atoms, are mandatory to observe the predicted interference phenomena. Only few experimental approaches towards these goals exist hitherto\cite{casabone_twoions_2015,reimann_twoatoms_2015}, and disclosing a realistic avenue for scaling towards many emitters is still a formidable challenge.

\begin{figure}
\includegraphics[width=88mm]{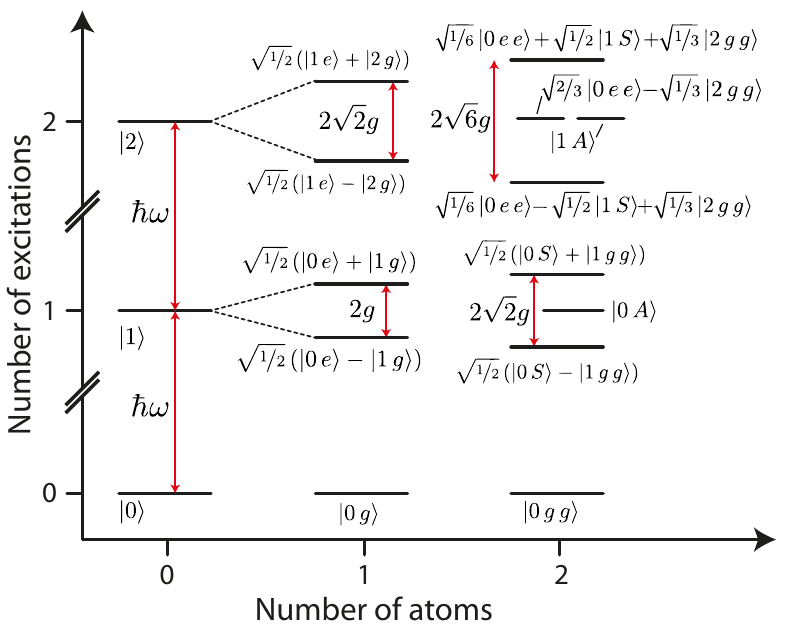}
\caption{
\textbf{Energy spectrum of the atom-cavity system.} For a single atom, the well-known anharmonic Jaynes-Cummings ladder is found. For two atoms, only a state that is symmetric with respect to exchange of the atoms $\ket{S}=(\ket{e\,g}+\ket{g\,e})/\sqrt{2}$ couples to the cavity, while the antisymmetric state $\ket{A}=(\ket{e\,g}-\ket{g\,e})/\sqrt{2}$ remains decoupled. In the singly-excited manifold, the symmetric state thus forms collective normal modes while $\ket{0\,A}$ remains at the unperturbed resonance energy $\hbar\,\omega$. In the doubly excited manifold for two atoms, all states that contain a single excitation in the cavity are detuned. Only the antisymmetric state $\ket{1\,A}$ and a superposition state with the cavity or the atoms being doubly excited remain at resonance energy $2\hbar\omega$.}
\label{fig:spectrum}
\end{figure}

Here we report on experiments with a deterministically prepared pair of $^{87}$Rb atoms permanently coupled to the mode of a $0.5\,$mm long optical Fabry-Perot resonator. Both atoms are driven with a laser propagating perpendicular to the cavity axis. Collective scattering of light into the cavity mode is studied as a function of the phase difference with which the atoms couple to both the cavity mode and the driving laser. The geometry resembles the proposed free-space situation\cite{richter_g2int_1991,schoen_bunching_2001,skornia_bunching_2001}, with the difference that the scattering phases are controlled by the atomic pattern instead of the angle of observation.

To tightly confine the atoms to well-defined positions, a two-dimensional optical lattice is implemented inside the cavity with periodicities of 386\,nm and 532\,nm along and perpendicular to the cavity axis, respectively (Fig.\,1a). Optical molasses cooling brings the atoms close to the mechanical ground state of the lattice. The residual spatial extent of $25\,$nm renders the atoms effectively point-like. During cooling, atomic fluorescence is collected by a high-numerical-aperture objective for imaging. Note that the close vicinity of the cavity mirrors (with a diameter of $1.5\,$mm) limits the achievable numerical aperture and poses a challenge in avoiding stray light. Nevertheless, with an exposure time of 0.75\,s it is possible to image individual atoms (Fig.\,1b) and obtain spot sizes with a full width at half maximum of 1.3\,\textmu m (see Supplementary Information). This allows us to find the origin of the emitted light with subwavelength precision by determining the center of Gaussian point-spread functions fitted to the images. Fig.\,1c shows a two-dimensional histogram of the difference vector of 45,000 atom pairs. We achieve unambiguous single-site detection of the difference vector between atoms.

An experimental run begins by loading a random distribution of atoms into the lattice. A fluorescence image of the atoms is automatically evaluated and excess atoms are identified. These are subsequently removed with a resonant push-out beam that is steered by an acousto-optical deflector and focused through the objective. In this way, a pair of atoms is deterministically prepared and data is taken until consecutively recorded atom images indicate the loss of one or both atoms.

The atom pair is prepared in a region of the intracavity lattice with a spatial extent so small that the modulus of the light-matter coupling strength $g$ can be regarded as constant (see Supplementary Information). The phase of $g$ alternates between $0$ and $\pi$ in adjacent trapping sites along the cavity (y-axis), while the phase of the running-wave driving laser changes continuously along the perpendicular x-axis. The difference phase $\phi$ for a pair of atoms is discrete due to the optical lattice, and therefore amounts to
$$\phi=\Delta n_x\cdot\frac{532}{780}\,2\pi+\Delta n_y \cdot \pi,$$
where $\Delta n_x$ and $\Delta n_y$ are the integer number of difference lattice sites along the transversal and the longitudinal direction, respectively, of the intracavity trap. Note that the single-site resolution is tantamount to identifying the exact value of the phase $\phi$.

\begin{figure}
\includegraphics[width=88mm]{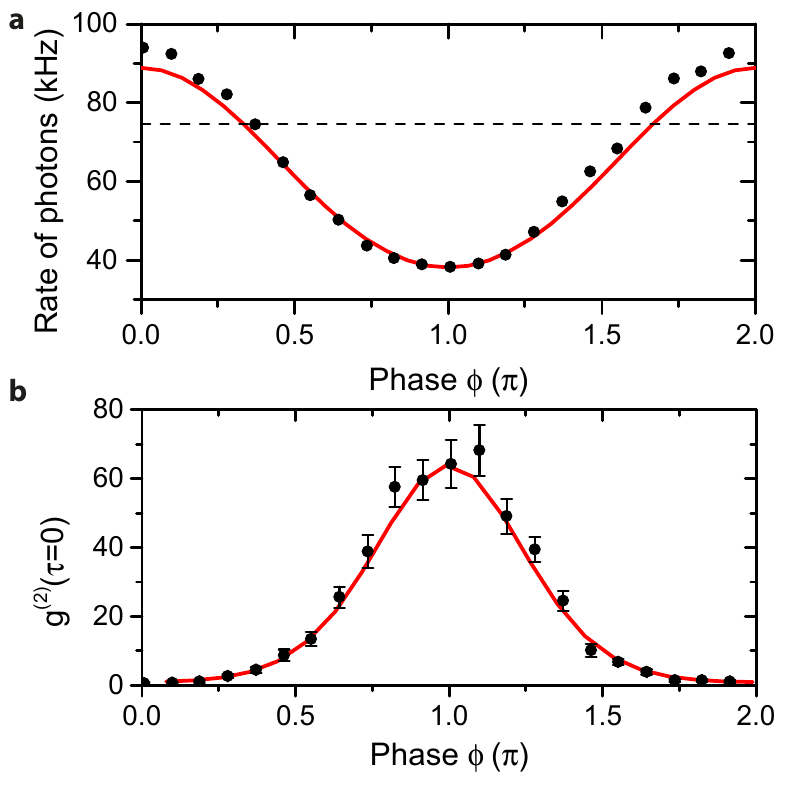}
\caption{\label{fig:intensity}
\textbf{Photon-emission rate (a) and \bm{$g^{(2)}(0)$} photon statistics (b) as a function of the relative phase \bm{$\phi$}.} \textbf{a}, For excitation resonant with the bare atom and the empty cavity at a Rabi frequency of $\Omega=920\,$kHz $\ll g$, the rate of photons emitted from the cavity shows a sinusoidal modulation as a function of the interatomic phase $\phi$. The dashed black line indicates the single-atom value of $75\,$kHz. Statistical error bars are within the symbols. \textbf{b}, The value of the second-order correlation function $g^{(2)}(0)$ indicating a transition from almost Poissonian to strongly super-Poissonian photon emission. The solid red lines in both plots are calculated from a numerical model without free parameters.}
\end{figure}

\begin{figure*}
\includegraphics[width=170mm]{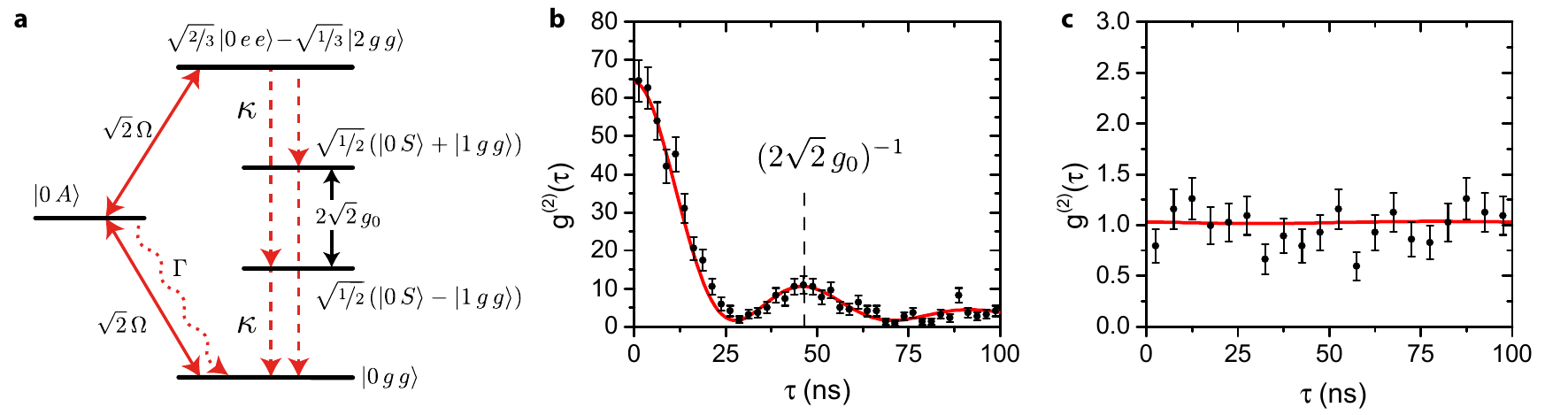}
\caption{\label{fig:g2}
\textbf{Emission dynamics for in- and out-of-phase coupling.} \textbf{a},  Mechanism of light scattering into the cavity mode in the out-of-phase configuration $\phi=\pi$: From the antisymmetric state $|0\,A\rangle$, the atom pair can either decay emitting a photon into free space at a rate $2\gamma$ or be further excited to a fully inverted state, which deposits a pair of photons in the cavity. \textbf{b}, Decay through the symmetric states leads to collective vacuum Rabi oscillations which are observable in the second-order correlation function. Consistently, we observe giant photon bunching $g^{(2)}(\tau=0)=64\pm 7$ with a revival after one period of the collective vacuum Rabi oscillation. \textbf{c}, For $\phi=0$, the data are compatible with the theoretical expectation of $g^{(2)}(\tau)\approx 1$ for an almost coherent intracavity field.}
\end{figure*}

Optical pumping initialises the atoms in the state $\ket{g}=\ket{\text{5S}_{\nicefrac[]{1}{2}}\text{,F=2,m}_F\text{=2}}$, and the cavity length is stabilised in such a way that a longitudinal mode red detuned from the intracavity lattice is resonant with the cycling transition $\ket{g}\leftrightarrow\ket{e}=\ket{\text{5P}_{\nicefrac[]{3}{2}}\text{,F=3,m}_F\text{=3}}$ at frequency $\omega$. On this transition, our parameters $(g,\kappa,\gamma)=2\pi\times(7.6,2.8,3.0)\,$MHz put us in the single-atom strong coupling regime of cavity quantum electrodynamics. Here, $\kappa$ and $\gamma$ are the decay rates of the cavity field and the atomic dipole, respectively. By polarizing the transversal driving beam orthogonal to the linear polarization of the trapping light, we can drive the $\ket{g}\leftrightarrow\ket{e}$ transition exclusively (see Supplementary Information). We thus realise the theoretical paradigm of an exactly known number of two-level atoms, two in our case, coupled with equal strength and known optical phases to the single light mode of a cavity and a transversal driving laser.

The energy eigenstates of the unperturbed Jaynes- and Tavis-Cummings Hamiltonians\cite{tavis_cummings_1968} are shown in Fig.\,2 up to the doubly excited manifold. For two (and more) atoms, all states have well defined parity with respect to exchange of atoms. While symmetric states $\ket{S}$ couple to the resonator, antisymmetric states $\ket{A}$ are decoupled and remain unshifted. The latter can be understood as a consequence of destructive interference of the emission from individual atoms. The driving laser induces transitions between the collective light-matter states, and the value of the phase difference $\phi$ determines which transitions can occur. For $\phi=0$ only transitions between states of equal atomic symmetry are possible, whereas in the case of $\phi=\pi$ every photon absorption changes the atomic symmetry. These different excitation pathways are expected to give rise to a strong dependency of the light emitted from the cavity on the phase $\phi$. In search for these interference effects, we record the emission of photons from the cavity with two single-photon detectors behind a 50:50 beam splitter. This setup allows us to measure both the intracavity photon number $\langle a^{\dagger}a\rangle$ and the photon number fluctuations via the second-order correlation function $g^{(2)}(\tau)=\langle a^{\dagger}a^{\dagger}(\tau)a(\tau)a\rangle/\langle a^{\dagger} a\rangle^2$, where $a$ ($a^{\dagger}$) are photonic annihilation (creation) operators.

Fig.\,3a shows the rate of photons emitted from the cavity as a function of $\phi$. The observed interference fringe displays several intriguing features which are not expected in a simple picture of two classical dipoles radiating into free space. The first one refers to the in-phase photon-emission rate (phase difference $\phi=0$) which is only a factor of $1.3$ larger than the rate observed for a single atom (measured dashed line in Fig.\,3a), not four-fold larger as one might predict for the coherent in-phase emission of two atoms. The second feature occurs for the out-of-phase situation (phase difference $\phi=\pi$) where the photon-emission rate drops below that of a single atom, but not even close to zero as intuitively expected. As shown in Fig.\,3b we furthermore observe an extremely strong phase-dependent effect in the normalised probability to simultaneously emit two photons, in accordance with predictions made for free-space interference\cite{schoen_bunching_2001,skornia_bunching_2001}. In fact, the value of the equal-time intensity-correlation function $g^{(2)}(\tau=0)$ rises from a value of $0.79\pm 0.16$ for in-phase coupling of the atoms ($\phi=0$) to $64\pm 7$ for out-of-phase coupling ($\phi=\pi$), indicating the transition from nearly Poissonian to strongly super-Poissonian photon emission. In the following we discuss these findings.

The weak dependency of the output power on the number of atoms for in-phase coupling has been predicted theoretically\cite{alsing_suppression_1992,zippilli_bragg_2004} for high atom-cavity cooperativity $C=g^2/2\gamma\kappa\gg 1$. A first indication was reported recently\cite{reimann_twoatoms_2015}. The effect is a consequence of the fact that in steady-state a field builds up inside the cavity that interferes destructively with the driving field at the positions of the atoms. In the limit of perfectly reflecting mirrors ($\kappa\rightarrow 0$), the driving and the cavity field are of equal strength and perfectly cancel each other at the position of the atoms. As a result, the intracavity power is independent of the number of atoms. For our cooperativity $C=3.4$, the steady state depends weakly on the number of atoms. The measured value is in agreement with a numerical simulation, the result of which is shown in Fig.\,3 and in the remainder of the paper as solid red lines. All parameters of this model were determined from independent single-atom experiments (see Supplementary Information).

When the phase between the atoms is nonzero, the driving field not only couples the ground state $|0\,g\,g\rangle$ to the symmetric normal modes $(\ket{0\,S}\pm\ket{1\,g\,g})/\sqrt{2}$ but also to the antisymmetric state $\ket{0\,A}$. For the particular case of $\phi=\pi$, the ground state is exclusively coupled to the state $\ket{0\,A}$ in the singly-excited manifold. In this state, the cavity is not populated, and decay back to the ground state via emission into free space competes with further excitation of the atoms to the doubly excited manifold (Fig.\,4a). Indeed, the state $\sqrt{\nicefrac[]{2}{3}}\ket{0\,e\,e}-\sqrt{\nicefrac[]{1}{3}}\ket{2\,g\,g}$ can be resonantly excited while $\ket{1\,A}$ remains unpopulated because it has the same symmetry as $\ket{0\,A}$. The two remaining states in the second manifold are detuned from the laser and therefore cannot be excited, too (Fig.\, 2). This two-step excitation path may thus lead to population of the cavity with a photon pair without admixture of single-photon components such that correlated emission is expected.

Fig.\,4b shows the $g^{(2)}(\tau)$ correlation function for out-of-phase coupling ($\phi=\pi$). Detection of a first photon from $\sqrt{\nicefrac[]{2}{3}}\ket{0\,e\,e}-\sqrt{\nicefrac[]{1}{3}}\ket{2\,g\,g}$ heralds preparation of the system in $\ket{1\,g\,g}$. The system is then in a superposition of the two normal modes and undergoes a collective vacuum Rabi oscillation. This manifests itself as a temporal oscillation in the $g^{(2)}$ correlation function. Consistent with the numerical simulation (red line), we observe a quick decay of the initial bunching peak followed by a revival at $(2\sqrt{2}\,g)^{-1}$, i.e. after a full period of the collective vacuum Rabi cycle. In contrast, $g^{(2)}(\tau)$ for the case of in-phase coupling (Fig.\,4c) is within statistical errors consistent with the expectation of an almost coherent intracavity field.

In summary, we demonstrated that the combination of deterministic preparation of an atom pair together with single-site-resolved imaging thereof enables us to enter the new regime of phase-resolved multi-atom cavity quantum electrodynamics. The experimental techniques presented here are mandatory requirements for future experiments aiming at the exploration of collective radiation effects predicted for multi-atom systems\cite{vogel_squeezing_1985,richter_g2int_1991,macovei_2007,fernandez_nonlinear_2007,gruenwald_entanglement_2010,habibian_quantumlight_2011,nikoghosyan_noon_2012,habibian_entanglement_2014} as well as the implementation of novel entanglement and quantum information processing schemes with multiple qubits\cite{pellizzari_quantcomp_1994,pachos_ioncavgate_2002,metz_quantumjumps_2006,kastoryano_dissent_2011}.

\begin{acknowledgments}
We thank A. Kochanke for the design of the objective, F. Saworski and C. Hahn for contributions in an early stage of the experiment and M. Uphoff for discussions. This work was supported by the European Union (Collaborative Project SIQS), by the Bundesministerium f\"ur Bildung und Forschung via IKT 2020 (QK\_QuOReP and Q.com-Q) and by the DFG via NIM.
\end{acknowledgments}

\clearpage
\renewcommand{\thefigure}{S\arabic{figure}}
\setcounter{figure}{0}

\section*{Supplementary Information}
\subsection*{Apparatus and loading of atoms}
Our system consists of a UHV chamber that holds the optical cavity and a rubidium dispenser, which directly loads a magneto-optical trap (MOT) $1\,$cm away from the cavity. The cavity is 498\,\textmu m long, leading to a free spectral range of $301\,$GHz, and supports a TEM$_{00}$ mode with a waist ($1/e^2$ intensity radius) of 30\,\textmu m. The finesse of the cavity is $55,000$, which results in a total field decay rate of the cavity of $\kappa=2\pi\cdot 2.8\,$MHz. The output coupling mirror has a transmission of $T_{\text{OC}}=100\,$ppm, while the second mirror has a transmission of only $T_2=4\,$ppm. This leads to a partial cavity field decay rate through the output coupling mirror of $\kappa_{\text{OC}}=2\pi\cdot2.4\,$MHz. Decay into a single free-space mode thus dominates the total cavity losses. The polarisation decay rate for $^{87}$Rb is $\gamma=2\pi\cdot 3.0\,$MHz and given the cavity mode volume $V$, the expected coherent coupling rate on the cycling transition $|5S_{\nicefrac[]{1}{2}},\text{F=2,m}_F\text{=}\pm2\rangle\leftrightarrow|5P_{\nicefrac[]{3}{2}},\text{F=3,m}_F\text{=}\pm 3\rangle$ with transition dipole moment $\mu_{ge}$ is
$$g=\sqrt{\frac{\omega}{2\epsilon_0V\hbar}}\,\mu_{ge}=2\pi\cdot 7.8\,\text{MHz}>(\gamma,\kappa),$$
which puts us in the strong coupling regime of cavity QED with a cooperativity $C=\frac{g^2}{2\kappa\gamma}=3.7$.

The atom-preparation sequence starts by loading the magneto-optical trap (MOT). Subsequently, a running-wave dipole trap at $1064\,$nm with its focal point halfway between the MOT and the cavity guides the atoms into the cavity after the MOT is switched off. Following a transfer time of $150\,$ms, which corresponds roughly to half a cycle of the longitudinal trap frequency, a standing-wave trap focused inside the cavity region is switched on. Atoms trapped in this $2.6\,$mK deep initial trap are cooled by application of a lin$\perp$lin optical molasses red detuned $22\,$MHz from the unshifted 5S$_{\nicefrac[]{1}{2}}$,F=2$\leftrightarrow$5P$_{\nicefrac[]{3}{2}}$,F=3 transition. Atomic fluorescence is imaged onto an \textit{Andor Ixon DU-897} electron multiplying CCD camera. During this stage, the blue and therefore repulsive intracavity lattice at a wavelength of $772.4\,$nm is held at a low power that is just sufficient to keep the cavity locked with a Pound-Drever-Hall technique but has no significant influence on the motion of the atoms. After image evaluation, a pushout beam is steered through the objective via an acousto-optical modulator to deterministically remove unwanted atoms. After preparation, the atomic pattern is positioned at the centre of the cavity mode by means of a tiltable glass-plate in the path of the 1064\,nm standing wave$^{\mathrm{S}1}$.

Subsequently, the intracavity lattice power is increased 20 fold to a trap depth of $0.8\,$mK while the red transversal lattice is lowered to a trap depth of $1.7\,$mK. The resulting two-dimensional lattice has degenerate trap frequencies along the x- and y-axis of $520\,$kHz. Along gravity (z-axis), the atoms are only weakly confined by the Gaussian envelope of the transversal lattice. Independent measurements show that the atoms are cooled to approximately 30\,\textmu K in the optical molasses, which is close to the mechanical ground state of the tightly confined directions.

When the intracavity lattice is switched on, atoms that were initially only weakly confined along the y-axis by the Gaussian envelope of the transversal lattice are pinned to random positions. The envelope of their positions in the two-dimensional lattice (Fig.\,1c) has a full width at half maximum (FWHM) of 3\,\textmu m. The waist of the red lattice is 16\,\textmu m such that the residual spread of the atoms results in an uncertainty of the light shift below $1\,\%$. The intracavity lattice populates a longitudinal mode whose frequency is $13$ free spectral ranges higher than that of the mode used for the described experiments. Therefore, a spatial beating pattern between the two modes with a periodicity of 38\,\textmu m appears. The intersection of the transversal optical lattice with the cavity mode is aligned such that the atoms are trapped in a region where the two longitudinal modes are exactly out of phase, such that the intracavity trap confines the atoms to antinodes of the mode at the $^{87}$Rb D$_2$ line. Given the small spatial extent of the trapping area in comparison to the beating patter, we can assume the modulus of the coupling $g$ to be constant. Limited by collisions with the background gas, the $1/e$-storage time for a single atom is $20\,$s, which is decreased by a factor of two for an atom pair.

\subsection*{Imaging in the 2D lattice}
During repetition of the actual experiment, interleaved with periods of molasses cooling, atomic fluorescence is collected via a specifically designed objective with a working distance of $24.9\,$mm through an $8\,$mm thick fused-silica vacuum window. The numerical aperture is limited along the x-axis by the entry aperture of the objective to $0.46$ and along the y-axis by the edges of the cavity mirrors to $0.32$. Collected fluorescence at $780\,$nm is imaged onto the EMCCD camera. The total system has a magnification of 0.48\textmu m/pixel and the minimum FWHM of an atom image limited by diffraction is $1.23\,$ and $1.78\,$pixel along the x- and y-axis, respectively.

\begin{figure}
\includegraphics[width=\columnwidth]{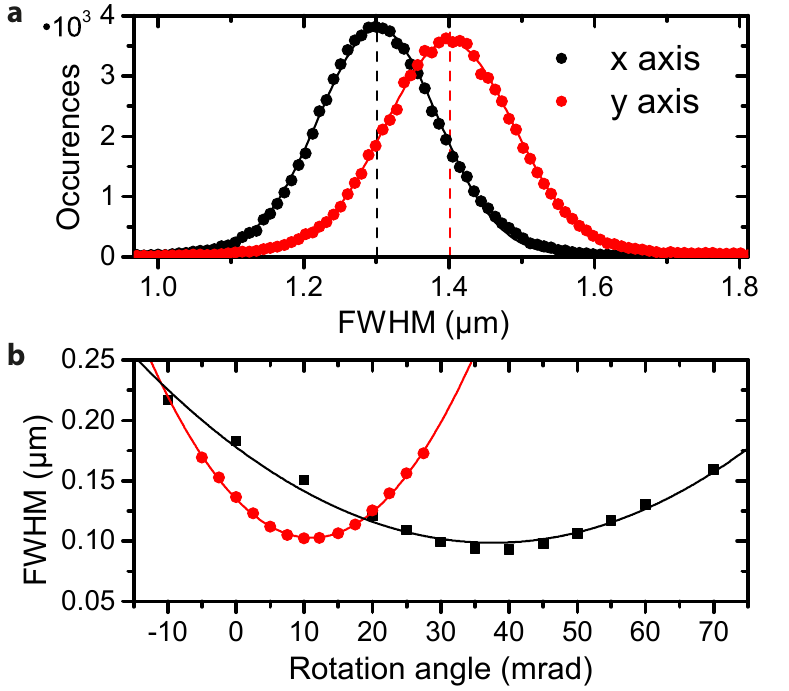}
\caption{\textbf{a}, Full width at half maximum (FWHM) of the fitted Gaussian point-spread functions along the two Cartesian directions. The solid lines are Gaussian fits. \textbf{b}, FWHM of the detected atom pairs' difference positions projected onto the two axes as a function of a global rotation applied to the data. The solid lines are quadratic fits to the data. For the two axes, minima are found at slightly different, nonzero rotation angles indicating a deviation of the lattice from orthogonality of $1.6^{\circ}$ and a global rotation of the lattice with respect to the camera's coordinate system of $0.64^{\circ}$.}
\label{fig_s1}
\end{figure}

The exposed images have a background signal of $450$ counts per pixel and a single atom accounts for approximately 18,000 additional counts. Gaussian point-spread functions are fitted to the atom images. In case of two atoms being too close, the data is discarded. Histograms of fitted widths along the x- and y-direction are shown in Fig.\,S1a. Along both directions, the diameter of the atom images is a factor $\approx 1.5$ greater than the theoretical diffraction limit. This discrepancy arises partly from the motion of the atoms along the weakly confined z-axis with an extent comparable to the focal depth of the objective. To quantify the achieved precision with which an atom's position can be determined and to characterise possible geometrical distortions of the lattice, we project the x- and y- difference coordinates of detected atom pairs, bin the data and evaluate the width of individual peaks in the periodic pattern (Fig.\,1c). Fig.\,S1b shows the FWHM of the peaks along the two Cartesian directions as a function of a global rotation applied to the coordinates. The two curves have minima at different, nonzero rotation angles indicative of a small non-orthogonality present in the lattice. To account for the global rotation and this non-orthogonal deformation, the following transformation is applied to the raw data:
$$\left(\begin{array}{c}x'\\y'\end{array}\right)=\left(\begin{array}{cc}C\beta & -S\beta\\0 &1\end{array}\right)\left(\begin{array}{cc}C\alpha &S\alpha\\-S\alpha &C\alpha\end{array}\right)\left(\begin{array}{c}x\\y\end{array}\right)$$
with ($C\beta$)$S\beta$ being the (co-)sine of the skew angle $\beta=1.6^{\circ}$ and ($C\alpha$)$S\alpha$ being the (co-)sine of the rotation angle $\alpha=0.64^{\circ}$. The minimum FWHM of $100\,$nm corresponds to a precision of $71\,$nm for the detection of a single atom and allows for basically unambiguous single-site detection. When the difference vector of an atom pair is found in terms of discrete trapping sites, the atom pair's measured difference vector is replaced by the exact value given by the above equation, which is then used to calculate the phase difference $\phi$. On average, approximately every $15\,$s an atom leaves its trapping site and is recooled into another site, thus performing a jump of up to several \textmu m. We detect these jumps on the images and discard all data that was taken during exposure of the image in which the jump appeared, the preceding image and the following image.

\subsection*{Excitation scheme and optical pumping}
The trapping light at a wavelength of $1064\,$nm, which is linearly polarised along the cavity axis, introduces a strong tensor light shift onto the excited $|$5P$_{\nicefrac[]{3}{2}}$,F=3$\rangle$ state. Fig.\,S2 shows the calculated energy structure at the trapping-light intensity of $1.1\cdot 10^{10}\,$Wm$^{-2}$, which is found via independent light shift spectroscopy$^{\mathrm{S}2}$. The degeneracy of the hyperfine Zeeman substates is lifted by up to $51\,$MHz. The cavity is not birefringent and when tuned into resonance with the cycling transition $|$5S$_{\nicefrac[]{1}{2}}$,F=2,m$_F$=2$\rangle\leftrightarrow|$5P$_{\nicefrac[]{3}{2}}$,F=3,m$_F$=3$\rangle$, a circularly polarised mode couples to the atom. The tensor light shift caused by the dipole trap shifts the $|$5P$_{\nicefrac[]{3}{2}}$,F=3,m$_F$=1$\rangle$ state by $46.4\,\text{MHz}$ out of resonance, which corresponds to more than 7 atomic linewidths. The orthogonal $\sigma^-$-polarised mode is therefore far detuned from the atom. The same shift allows us to couple exclusively to this circularly polarised atomic dipole with linearly polarised light impinging orthogonally to the cavity axis. In the atomic reference frame, this light drives $\sigma^+$ and $\sigma^-$ transitions with equal strength. But again, the $\sigma^-$-polarised transition is far detuned, such that we can treat the atoms as effective two-level systems with a well known atomic dipole, which allows for quantitative predictions.

\begin{figure}
\includegraphics[width=\columnwidth]{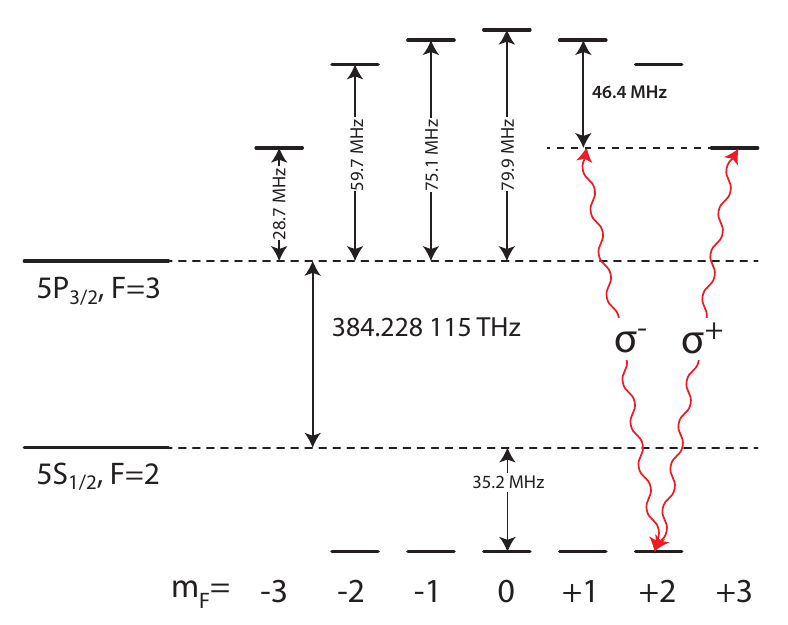}
\caption{Light shifts of the excited energy eigenstates caused by the $1064\,$nm dipole trap (not to scale). A tensor light shift of the excited 5P$_{\nicefrac[]{3}{2}}$,F=3 state shifts the transition from 5S$_{\nicefrac[]{1}{2}}$,F=2,m$_F$=2 to 5P$_{\nicefrac[]{3}{2}}$F$'$=3,m$_F$=1 out of resonance by more than $7$ atomic linewidths.}
\label{fig_s2}
\end{figure}

In order to prepare all atoms in the $|$5S$_{\nicefrac[]{1}{2}}$,F=2,m$_F$=+2$\rangle$ state, circular pumping light resonant with the cavity is applied for 80\,\textmu s in combination with a weak repumping beam incident perpendicular to the cavity that brings atoms back from the F=1 ground state. A problem appears when trying to optically pump more than a single atom while the cavity is held on resonance with the cycling transition. The first atom that couples to the cavity will lead to a normal-mode splitting of the cavity resonance and suppress the intracavity intensity by a factor of $(1+2C)^2=69$, thereby effectively switching off the light and preventing the second atom from being pumped. To circumvent this problem, we lower the intensity of the dipole trap during optical pumping by $20\,\%$. Besides detuning the cycling transition from the cavity resonance, the above mentioned tensor light shifts on the other $m_F$ states are reduced as well, which simplifies optical pumping. The first atom to appear on the cycling transition now only leads to a relatively small attenuation of the pumping light by a dispersive shift of the cavity such that all atoms can be optically pumped. The efficiency of optical pumping is characterised by performing cavity-reflection spectroscopy and we find a single-atom efficiency of $\eta=0.87$.

\subsection*{Theoretical model}

\begin{figure}
\includegraphics[width=\columnwidth]{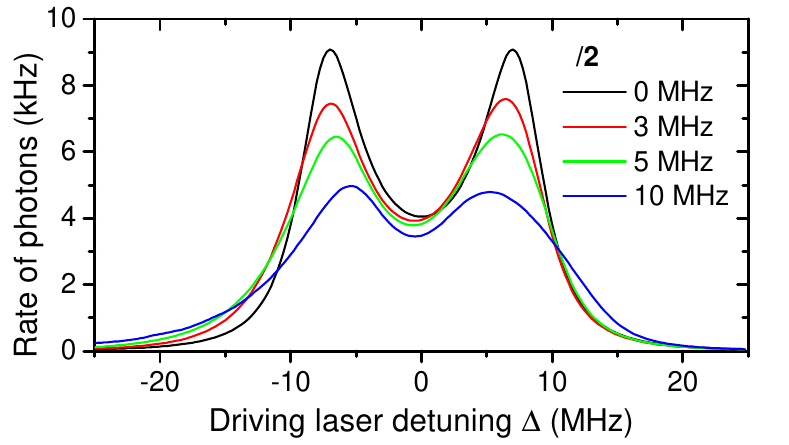}
\caption{Simulated photon emission rate for our cavity QED parameters $(g,\gamma,\kappa)=2\pi\cdot(7.6,3.0,2.8)\,$MHz and a driving Rabi frequency of $\Omega=2\pi\cdot 200\,$kHz for different values of the temperature parameter $\tau$ and a  detuning between an atom at the trap bottom and the cavity that is equal to $\tau$.}
\label{fig_s3}
\end{figure}

The single cavity mode is described by the bosonic annihilation (creation) operator $a$ ($a^{\dagger}$) and the two level atoms with ground state $|g\rangle$ and excited state $|e\rangle$ are described by the atomic operators $\sigma=|g\rangle\langle e|$ and $\sigma^+=|e\rangle\langle g|$. For a single atom in the cavity, the master equation of the driven system is
\begin{eqnarray*}
H &=& -\Delta_c a^{\dagger}a-\Delta_a\sigma^+\sigma^-+g(a\sigma^++a^{\dagger}\sigma^-)+H_{\text{drive}}\\
H_{\text{drive}} &=& \frac{\Omega}{2}(\sigma^++\sigma^-)\\
\frac{d}{dt}\rho &=& -i[H,\rho]\\
&+&\gamma (2\sigma^-\rho\sigma^+-\sigma^+\sigma^-\rho-\rho\sigma^+\sigma^-)\\
&+&\kappa (2a\rho a^{\dagger}-a^{\dagger}a\rho-\rho a^{\dagger}a)
\end{eqnarray*}
where $\Delta_c$ and $\Delta_a$ are the detuning of the cavity and atomic transition with respect to the driving laser frequency respectively. The two collapse operators describe decay of the cavity field and of atomic polarisation respectively. We numerically find the steady state $\rho_{\text{ss}}$ of the system and calculate the expected rate of photons as $R=30.4\,\text{MHz}\cdot\text{Tr}(a^{\dagger}a\,\rho_{\text{ss}})$. In order to model inhomogeneous broadening due to finite temperature of the trapped atom, we average over Boltzmann-distributed atomic detunings. Assuming classical motion and all degrees of freedom to be thermalised, the following integral is found for the expectation value:
$$\langle R(\Delta_a)\rangle=\frac{4}{\sqrt{\pi}}\int_0^{\infty}R(\Delta_a+\tau\,r^2)r^2e^{-r^2}dr,$$
where $\tau=\frac{\alpha_e-\alpha_g}{\hbar\alpha_g}k_BT$ is a temperature parameter normalised to the relative difference in dynamic polarisability of the excited ($\alpha_e$) and ground state ($\alpha_g$). The assumptions made in the derivation of this formula are only partially valid. See ref.$\mathrm{S}2$ for a discussion of validity. When the cavity is tuned such that it is resonant with an atom at the bottom of the trap that experiences the maximum light shift, the effect of a finite temperature is an asymmetric broadening towards lower transition frequencies as the atom explores spatial regions with lower trap intensity. Choosing a slightly higher power such that an atom at the bottom of the trap is blue detuned from the cavity restores a symmetric normal-mode spectrum. Fig.\,S3 illustrates the residual effect of a finite temperature when the bottom of the trap is detuned by the value $\tau$. The predominant qualitative change is a loss of contrast of the observed curve. For temperatures exceeding $\tau/2\pi\approx 5\,$MHz, a visible asymmetry in the form of a broadening of the blue resonance with respect to the red becomes apparent.

\begin{figure}
\includegraphics[width=\columnwidth]{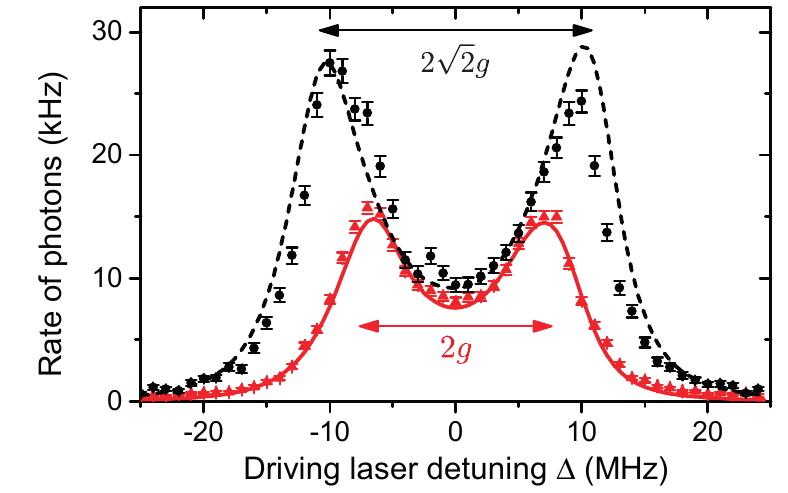}
\caption{Cavity emission as a function of the detuning of the driving laser in case of a single (red triangles) and two atoms (black dots) that couple in phase ($\phi=0$). The Rabi frequency is $\Omega=2\pi\cdot 300\,$kHz. The solid red line is a fit to the data and the dashed black line is a numerical calculation for the case of two atoms using the paramaters found by fitting to the single-atom data. The collectively-enhanced coupling strength $g_{\text{eff}}=\sqrt{2}g$ manifests itself in a greater separation of the two resonances while the amount of emission on the bare atoms' resonance increases only slightly.}
\label{fig_s4}
\end{figure}

Fig.\,S4 shows experimental data taken with a single atom (red triangles) and with two atoms that couple in phase (black dots). The solid red line is a fit of the described numerical model extended to take the non-unity optical pumping efficiency into account. As all possible transitions are detuned by several linewidth, when optical pumping did not succeed to initialise the atom in $|F=2,m_F=+2\rangle$ (Fig.\,S2), we assume that no light is created in these cases. The pumping efficiency $\eta$ thus only rescales the y-axis. The validity of this model is further substantiated by measuring the polarisation of the light emitted from the cavity. We find the light to be right-circularly polarised with an admixture of orthogonally polarised light $<1\,\%$ which is close to the attenuation ratio of the involved polarisation optics. For the dataset shown in Fig.\,S4, we find the atomic detuning with respect to the cavity to be $2.89\,$MHz and the temperature parameter $\tau/2\pi=2.28\,$MHz.

For a pair of atoms with identical light-matter coupling strength $g$, the master equation of the system is extended by collapse operators to describe free-space emission by the second atom and coupling terms to the quantised cavity field and the classical driving field. By means of gauge transformations $e^{i\varphi}a\rightarrow \tilde{a}$, the phases appearing for the different couplings are collected in front of one term. Expectation values of intensities $\tilde{a}^{\dagger}\tilde{a}$ and fluctuations thereof are left unaffected by these transformations and we find the following Hamiltonian:
\begin{eqnarray*}
H&=&-\Delta_c a^{\dagger}a-\Delta_{a_1}\sigma_1^+\sigma_1^--\Delta_{a_2}\sigma_2^+\sigma_2^-\\
&+&g\left[(\sigma_1^++\sigma_2^+)a+(\sigma_1^-+\sigma_2^-)a^{\dagger}\right]\\
&+&\frac{\Omega}{2}(\sigma_1^++\sigma_1^-+e^{i\phi}\sigma_2^++e^{-i\phi}\sigma_2^-)
\end{eqnarray*}

The effects of imperfect optical pumping and thermal averaging are taken into account in the same way as described above for a single atom assuming identical atomic detunings $\Delta_a$, uncorrelated optical pumping with single-atom efficiency $\eta$ and thermal motion described by $\tau$. The emission rate from the cavity is found to be $\langle R_2\rangle=\eta^2\langle R_2'\rangle+2\eta(1-\eta)\langle R_1\rangle$, where $R_2'$  and $R_1$ are thermally averaged emission rates for two and a single atom, respectively. The dashed black curve in Fig.\,S4 is calculated applying this numerical model for the case of $\phi=0$ and agrees well with the data.

\begin{figure}[!t]
\includegraphics[width=\columnwidth]{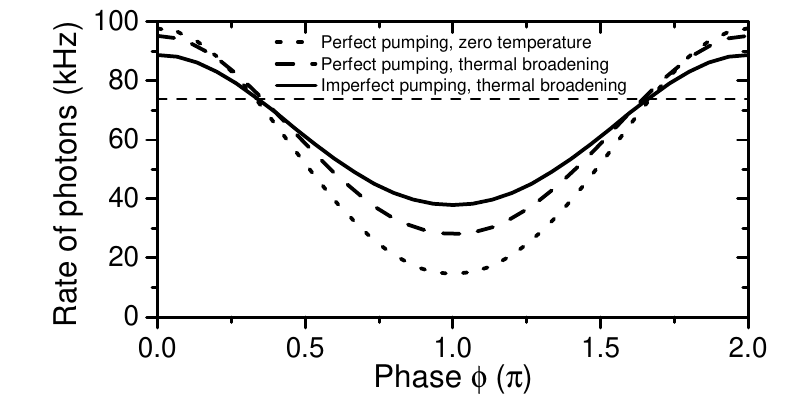}
\caption{Simulated output photon rate as a function of the interatomic phase $\phi$ including different experimental imperfections. The horizontal line indicates the single-atom value of $75\,$kHz.}
\label{fig_s5}
\end{figure}

Fig.\,S5 shows the different contributions of detrimental experimental imperfections to the model used to calculate the solid red line in Fig.\,3(a). The dotted line in Fig.\,S5 is calculated without thermal averaging and assuming perfect initial state preparation of the atom pair. Even under these perfect conditions, the observed intensity is only reduced by a factor of five for destructive interference of the atoms, because atomic saturation and emission of photon pairs contribute a significant part of the light. Taking thermal broadening into account increases the intensity at $\phi=\pi$ by a factor of 2 as inhomogeneous broadening reduces the indistinguishability of the atoms. An additional quarter of the observed light stems from single-atom emission in cases of imperfect optical pumping.

\section*{References for Supplementary Information}
\begin{enumerate}[S1.]
\setlength{\itemsep}{-\parsep}
\item
Nu{\ss}mann, S., Hijlkema, M., Weber, B., Rohde, F., Rempe, G. \& Kuhn, A. Submicron positioning of single atoms in a microcavity. \textit{Phys. Rev. Lett.} \textbf{95}, 173602 (2005).
\item
Neuzner, A., K{\"o}rber, M., D{\"u}rr, S., Rempe, G. \& Ritter, S. Breakdown of atomic hyperfine coupling in a deep optical-dipole trap. \textit{Phys. Rev. A} \textbf{92}, 053842 (2015).	
\end{enumerate}
\clearpage
\end{document}